\documentclass{aastex631}
\usepackage{newtxtext,newtxmath}

\begin{document}

\title{Magneto hydrodynamic simulations of the supernova remnant G1.9+0.3}

\author[0000-0002-0829-7553]{Shaobo Zhang}
\affiliation{National Astronomical Observatories, Chinese Academy of Sciences, Beijing 100012, People's Republic of China}
\affiliation{School of Astronomy and Space Science, University of Chinese Academy of Sciences, Beijing 100049, People's Republic of China}

\author[0000-0002-9079-7556]{Wenwu Tian}
\affiliation{National Astronomical Observatories, Chinese Academy of Sciences, Beijing 100012, People's Republic of China}
\affiliation{School of Astronomy and Space Science, University of Chinese Academy of Sciences, Beijing 100049, People's Republic of China}

\author[0000-0001-8261-3254]{Mengfei Zhang}
\affiliation{School of Physics, Zhejiang University, Zhejiang 310058, China}

\author[0000-0003-3775-3770]{Hui Zhu}
\affiliation{National Astronomical Observatories, Chinese Academy of Sciences, Beijing 100012, People's Republic of China}

\author[0000-0002-6322-7582]{Xiaohong Cui}
\affiliation{National Astronomical Observatories, Chinese Academy of Sciences, Beijing 100012, People's Republic of China}




\begin{abstract}

The youngest Galactic supernova remnant G1.9+0.3 shows a discrete feature between its radio and X-ray morphologies. The observed radio morphology features a single maximum in the north, while the X-ray observation shows two opposite 'ears' on the east and west sides. Using 3D magneto hydrodynamical simulations, we investigate the formation of the discrete feature of the remnant. We have tested different parameters for better simulation and reproduced similar discrete features under an environment with density gradient and an environment with clump, which provides a possible explanation of the observation.

\end{abstract}



\section{Introduction} \label{sec:intro}

A supernova remnant (SNR) is the structure resulting from the interaction between supernova ejecta and the interstellar medium. SNRs play an important role in many astrophysical processes, such as cosmic-ray acceleration and dust formation, but many details about the evolution of SNR are still unclear, especially in its earlier phase. SNR G1.9+0.3 is believed to be the youngest galactic supernova remnant with an age of $\sim$ 150 yr \citep{2020MNRAS.492.2606L} and could be a potential cosmic-ray accelerator. This young SNR provides us a precious sample to understand the physics of SNRs, especially for the early stages of their evolution, and the particle acceleration.

\begin{figure}
\gridline{\fig{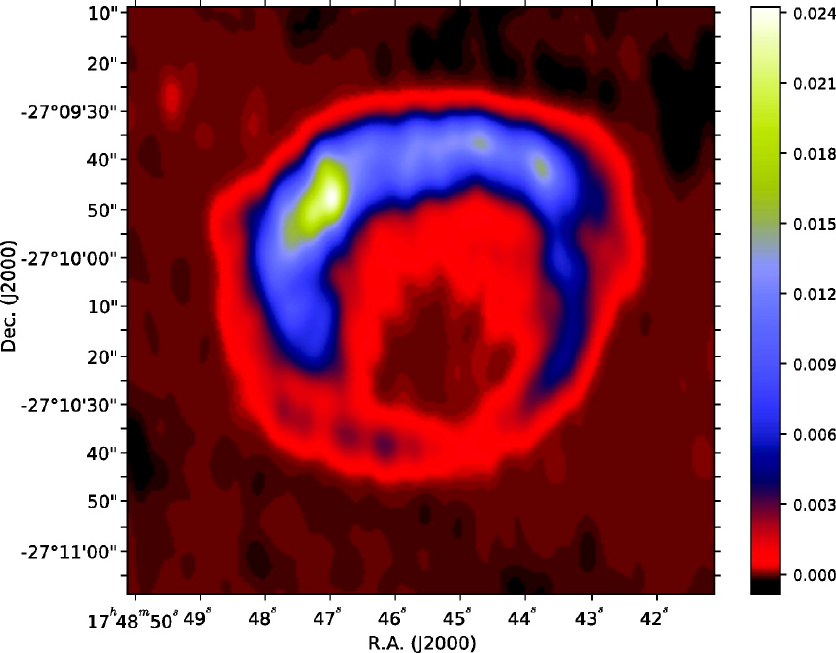}{0.49\textwidth}{(a)}
          \fig{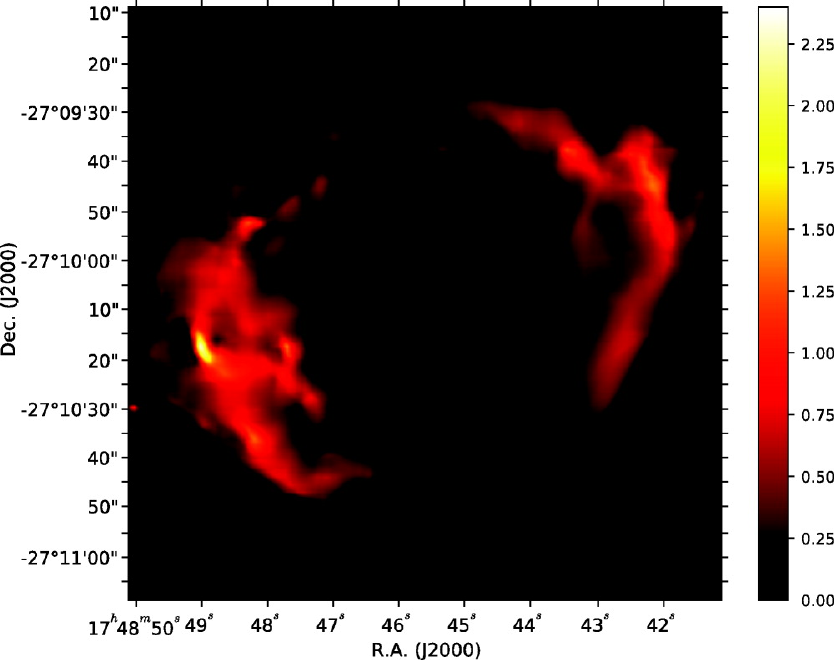}{0.49\textwidth}{(b)}}
\caption{The observed morphology of SNR G1.9+0.3. (a) 1.5 GHz VLA image, (b)1.5 - 6 keV Chandra image. Both images are from \cite{2008ApJ...680L..41R}.}
\label{Fig1}
\end{figure}

G1.9+0.3 was first detected by \cite{1984Natur.312..527G} and  has a shell-like morphology at radio band. \cite{2008ApJ...680L..41R} observed the remnant with Chandra for 50 ks, and placed it at a distance of about 8.5 kpc based on the X-ray absorption, which is consistent with a distance measurement by analyzing spectrum of H I and CO towards to the remnant \citep{2020MNRAS.492.2606L}. G1.9+0.3 has a radio feature different from its X-ray feature, as depicted in Figure \ref{Fig1}. The observed radio morphology is asymmetric, featuring a bright half ring to the north. In the X-ray band, the remnant shows bilateral structures, with two opposite 'ears' on the east and west sides.

The morphology and evolution of an SNR depend heavily on the properties of the progenitor and the surrounding interstellar medium, while we always know little  about them. Numerical simulations provide an effective method to study the SNR evolution and have been widely used with the improvement of computational power. \cite{2014MNRAS.437..898T} performed 3D numerical simulations to reproduce the asymmetric shape of Kepler’s supernova remnant in X-rays. \cite{2017ApJ...849..147Z}  identified a new northeast edge for SNR W51C based on simulation results. \cite{2015MNRAS.450.1399T} modeled SNR G1.9+0.3 inside a planetary nebula. Here, we will also use numerical simulations to study SNR G1.9+0.3 under two different environments.

In this paper, we focus on the discrete feature between X-ray and Radio morphology of G1.9+0.3 through the numerical simulation. We perform 3D magneto-hydrodynamical (MHD) simulations of SNR G1.9+0.3 and synthesize the emission map based on the simulation results. The paper is structured as follows. In Sect.2, we will describe the simulation setup and list the initial parameters. In Sect.3, we will show the simulation results of G1.9+0.3. {We will discuss the limitations of this work in Sect.4, and give our summary in Sect.5.}


\section{Numerical setup} \label{sec:setup}

We perform the simulations using the high-resolution multidimensional hydrodynamic code PLUTO (Version 4.2;\cite{2007ApJS..170..228M,2012ApJS..198....7M}). The code solves the system of MHD conservation laws by using a cell-centered finite-volume approach and achieves highly scalable parallel performance through the message passing interface standard (MPI). The simulation satisfies the conservation equations as follows:
\begin{equation}\label{eq1}
  \left\{\begin{array}{l}
  \frac{\partial \rho}{\partial t}+ \nabla\cdot(\rho\boldsymbol{v})=0
 \\\frac{\partial \rho \boldsymbol{v}}{\partial t}+ \nabla\cdot(\rho\boldsymbol{vv}-\boldsymbol{BB})+\nabla\cdot P^{*}=0
 \\\frac{\partial E}{\partial t}+ \nabla\cdot[(E+P^{*})\boldsymbol{v}-\boldsymbol{B}(\boldsymbol{v}\cdot \boldsymbol{B})]=0
 \\\frac{\partial \boldsymbol{B}}{\partial t}+ \nabla\times(\boldsymbol{v}\times\boldsymbol{B})=0
\end{array}\right.
\end{equation}
where $\rho$ is the matter density, $t$ is the time, $\boldsymbol{v}$ is the matter velocity, and $\boldsymbol{B}$ is the magnetic field intensity.
\begin{equation*}
  P^{*}=P+\frac{\boldsymbol{B}^{2}}{2}, E=\frac{P}{\gamma-1}+\frac{\rho\boldsymbol{v}^{2}}{2}+\frac{\boldsymbol{B}^{2}}{2}
\end{equation*}
are the total pressure(gas pressure $P$ and and magnetic pressure) and the total energy density, in which the adiabatic index $\gamma = 5/3$.

The classification of G1.9+0.3 is still unclear, but the high expansion speed and the Fe K emission support a thermonuclear origin \citep{2013ApJ...771L...9B}. In our simulations, G1.9+0.3 is assumed to be a Type Ia supernova remnant. we take the ejected mass as 1.4 $M_{\odot}$ and the explosion energy as $1\times10^{51}\,\rm{erg}$. We set the initial radius as 0.3 pc and the corresponding initial time as 11 years. At the initial time, G1.9+0.3 is in the free expanding stage, and the ejecta has a power-law density profile with a uniform inner region:
\begin{equation}\label{eq2}
  \rho_{ej}(r,t)=
  \left\{\begin{array}{ll}
       \rho_{c}(t) & r < r_{c}\\
       \rho_{c}(t)(r/r_{c})^{-n} & r\ge r_{c}
\end{array}\right.
\end{equation}
in which $\rho_{c}$ is the density of the inner region,$r_{c}$ is the radius of the inner region. We take power law index n=7, which is a classical value for a Type Ia supernova remnant(e.g \citealt{1981ApJ...246..267C}).

\begin{table}
\begin{center}
\caption{Summary of Simulation Parameters}\label{Tab1}
 \begin{tabular}{lll}
  \hline\noalign{\smallskip}
Parameters &  Value  & References                    \\
  \hline\noalign{\smallskip}
Ejecta mass              & 1.4 $\it{M}_\odot$  &1 \\
Initial explosion energy & $1\times 10^{51}$ $\rm{erg}$  &1 \\
Initial radius             & 0.3 pc &... \\
Distance                 & 8.5 kpc  &2 \\
\hline 
Mean density             & 0.21 $\rm{cm^{-3}}$  &... \\
Density Gradient         & 0.8 $\rm{cm^{-3}\,pc^{-1}}$  &... \\
Magnetic field intensity & 1 $\rm{\mu G}$&3 \\
Electron Spectral Index        & 2.3   &4,5 \\ \hline
\end{tabular}
\tablerefs{ (1)\cite{2020pesr.book.....V}; (2)\cite{2020MNRAS.492.2606L}; (3)\cite{Haverkorn2015}; (4)\cite{1999ApJ...525..368R}; (5)\cite{2008ApJ...680L..41R};}
\end{center}
\end{table}

To reproduce the observed morphology, we consider two scenarios of the ambient medium in our simulations. In scenario A, we perform the simulation in an environment with density gradients, whose number density distribution is as follows: \citep{2020RAA....20..154Y}:
\begin{equation}\label{eq3}
  n(r)= \left\{\begin{array}{ll} n_{0}+k\boldsymbol{\xi}\cdot\boldsymbol{r} & \boldsymbol{\xi}\cdot\boldsymbol{r} > r_{0}\\n_{0} & \boldsymbol{\xi}\cdot\boldsymbol{r}\le r_{0}
\end{array}\right.
\end{equation}
\begin{equation}\label{eq4}
\boldsymbol{\xi}=\sin\theta\cos\phi\boldsymbol{e_x}+\sin\theta\sin\phi\boldsymbol{e_y}+\cos\theta\boldsymbol{e_z}
\end{equation}
{in which $\boldsymbol{\xi}$ is the unit vector in the direction of the density gradient, and parallel to the xy-plane with $\theta=120^{\circ},\phi=0^{\circ}$ in this work}. We set the density gradient $k=0.8 \,\rm{{cm}^{-3}\,{pc}^{-1}}$, the background number density $n_{0}=0.21\,\rm{{cm}^{-3}}$ and the judgment condition $r_{0}=1.5\,\rm{pc}$. In scenario B, we insert an ellipsoidal clump in the northeast region. We take the number density of the clump as $10 \,\rm{{cm}^{-3}}$, and the distance from the center of the clump to the origin is $1.875\,\rm{pc}$. As there is no information on the initial structure of the magnetic field, we set a uniform magnetic field of 1 $\rm{\mu G}$ \citep{Haverkorn2015}, and the direction is identical to the density gradient.

To compare with the observation, we need to synthesize the relative flux in radio and X-rays. The emission of G1.9+0.3 in both bands is primarily synchrotron-based \citep{2009ApJ...695L.149R,2020MNRAS.492.2606L}, thus we only consider the contribution from synchrotron processes. To obtain the relative flux, we assume that the electrons have a power-law distribution with a high-energy exponential cutoff and the number density of electrons is proportional to the matter density. We set the power-law spectral index as 2.3 and the cutoff energy $E_{\rm{cut}}=58(v_s/v_0)\,\rm{TeV}$, in which $v_s$ is the local shock velocity and $v_0=1300 \,\rm{km/s}$ is the observed shock velocity of G1.9+0.3 \citep{1999ApJ...525..368R,2009ApJ...695L.149R,2011ApJ...737L..22C}

Our simulations are based on a 3D MHD grid in Cartesian (x, y, z) geometry that extends 6 pc in all directions. We employ a uniform grid of 400×400×400, with a corresponding spatial resolution of $1.5\times10^{-2}\,\rm{pc\,{pixel}^{-1}}$. Our simulations cover the evolution of the remnant over 11 to 158 years. Gravitation and radiation cooling have little influence on the simulation, so we ignore these effects. The simulation's parameters are summarised in Table 1.

\section{Results} \label{results}

\begin{figure}
\gridline{\fig{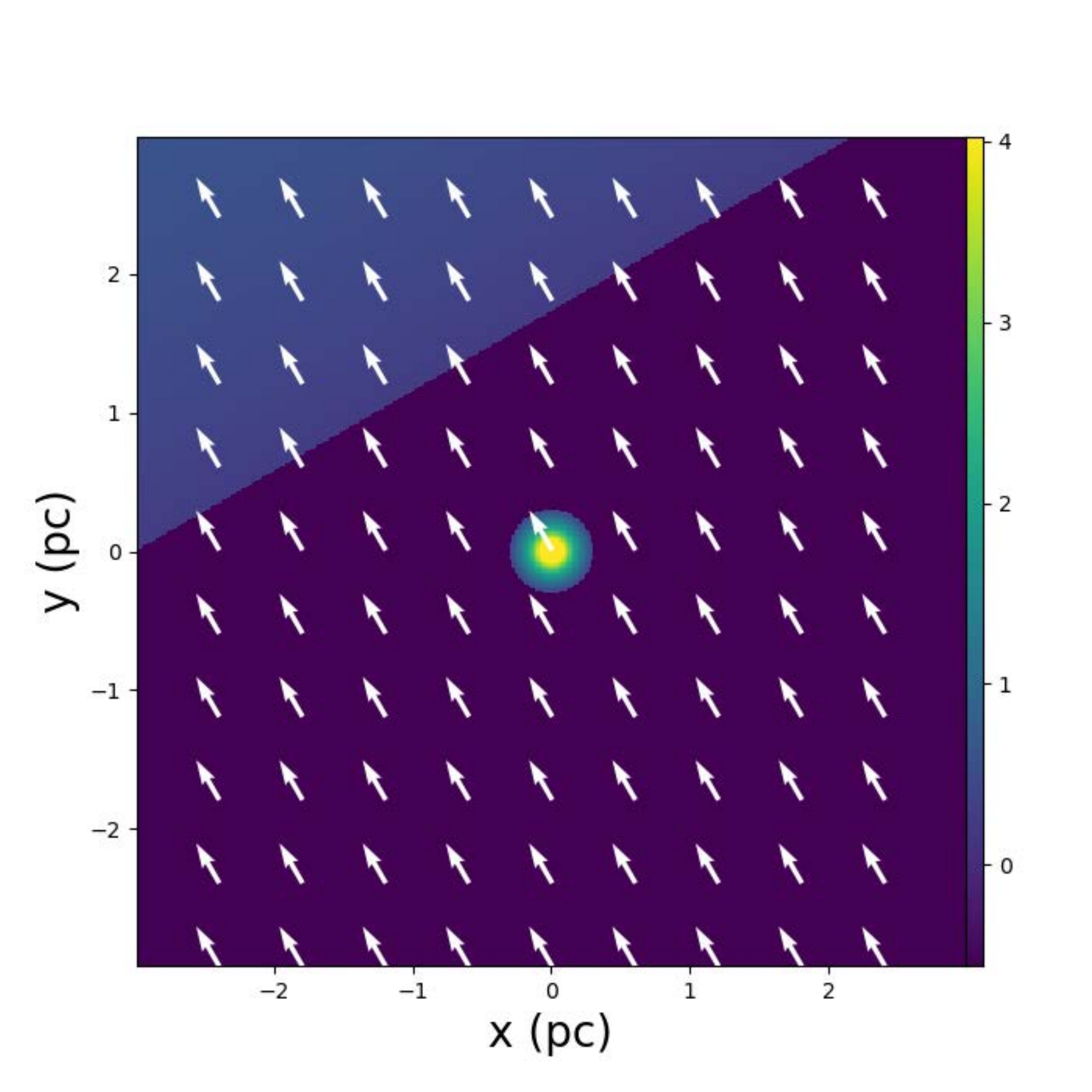}{0.49\textwidth}{(a)}
          \fig{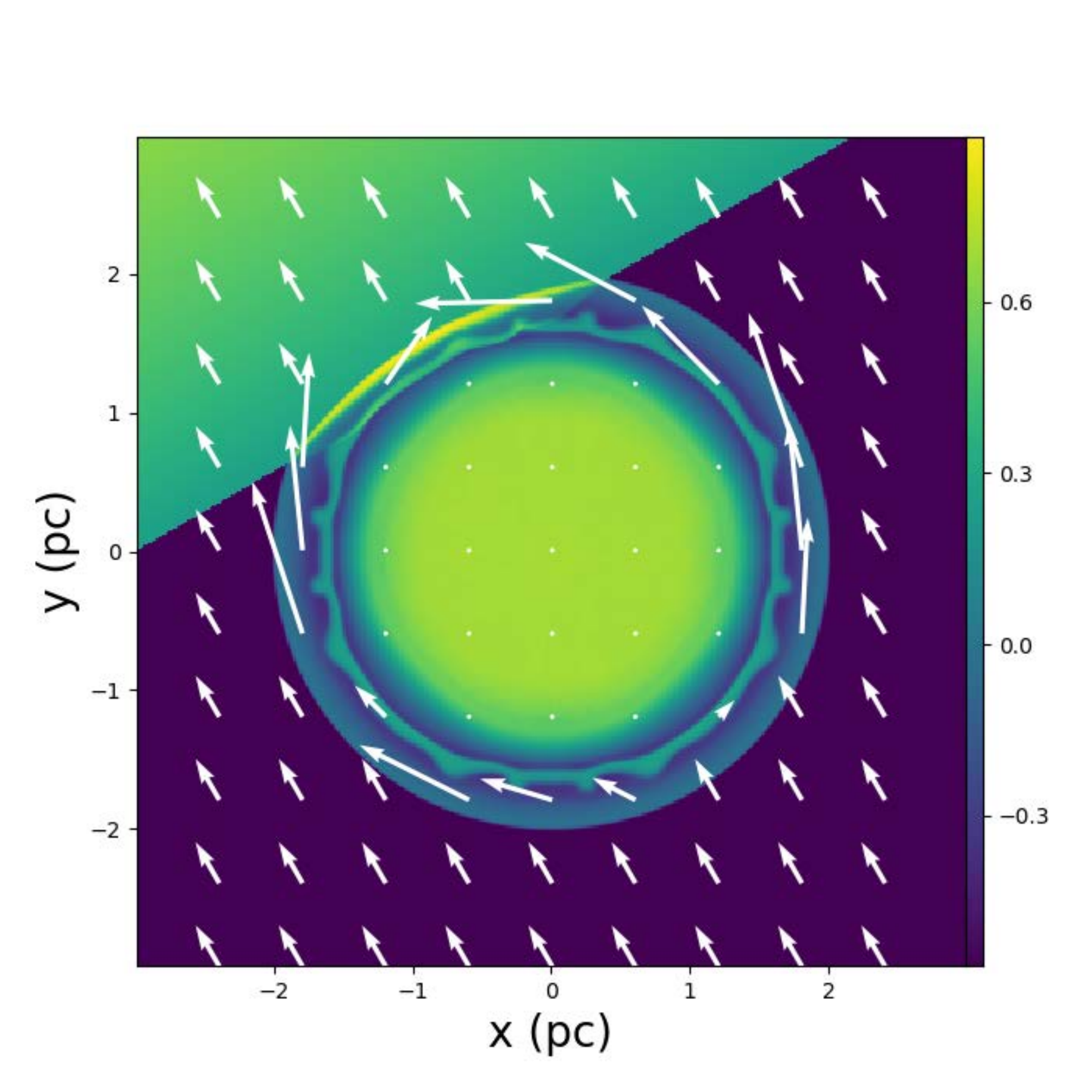}{0.49\textwidth}{(b)}}
\gridline{\fig{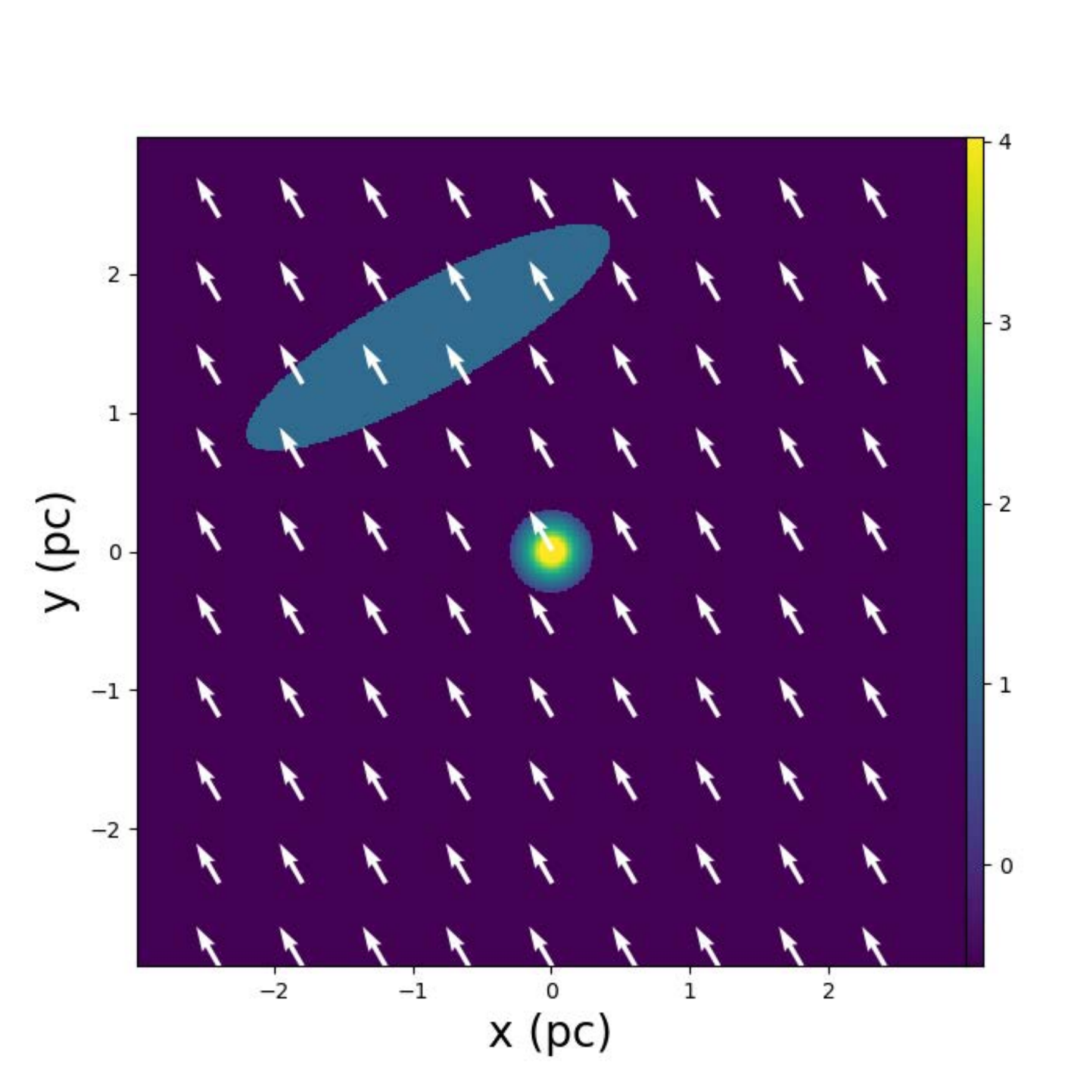}{0.49\textwidth}{(c)}
          \fig{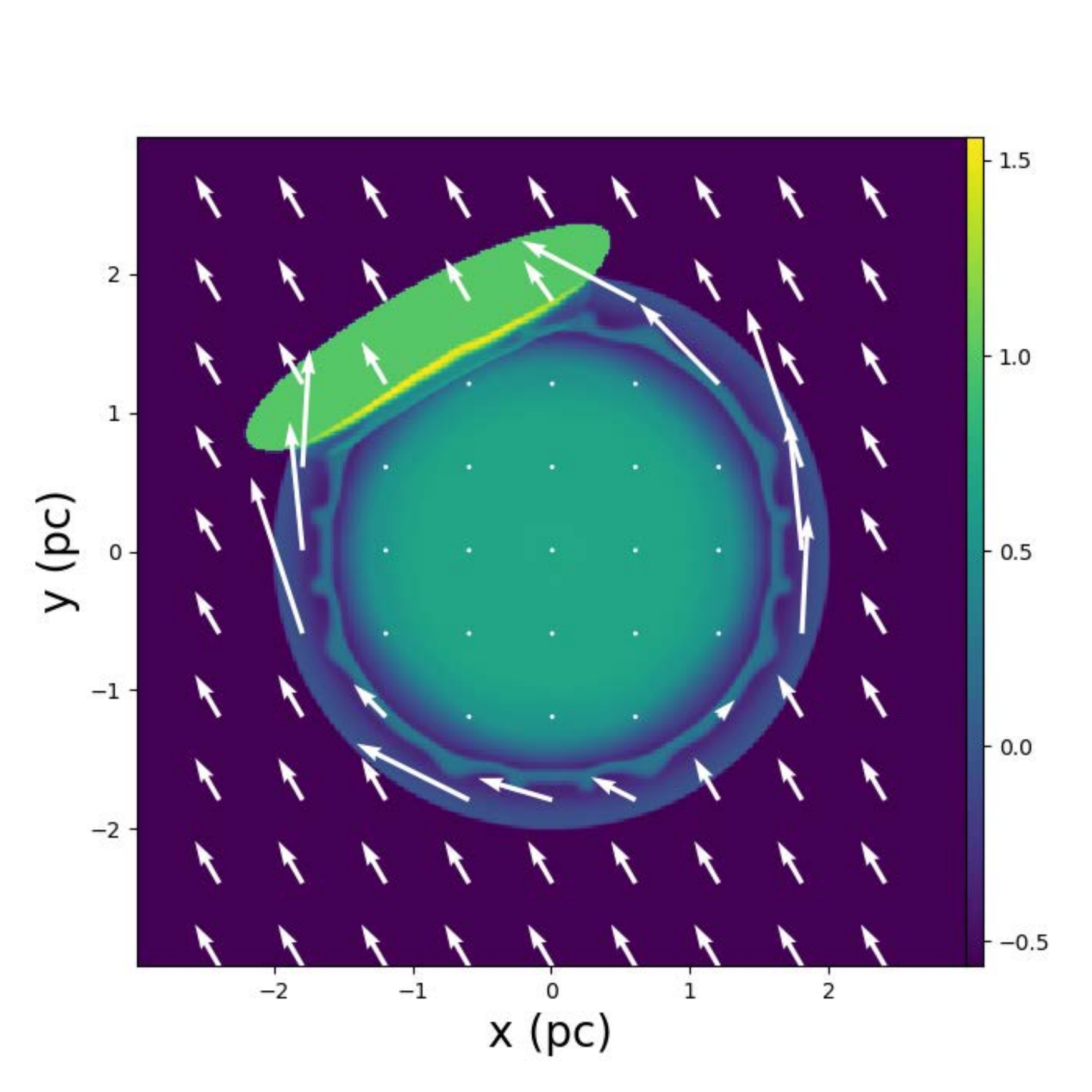}{0.49\textwidth}{(d)}}
\caption{The density–magnetic field images with different times. {The background is the numerical density distribution. The white arrows indicate the direction of the magnetic field, and the arrow length is proportional to the magnetic field strength.} Panels (a) and (b) correspond to scenario A, and panels (c) and (d) to scenario B. The left two pictures show the result at the start time, and the right two pictures show the result at the age of 148 years.}
\label{Fig2}
\end{figure}

In this section, we show the MHD simulation results and the synthesized synchrotron emission maps for both scenarios. {All figures are depicted under the assumption that the observer's line of sight is perpendicular to the xy-plane. For the sake of description, we assume that the positive direction of the x-axis is westward and the positive direction of the y-axis is northward}. We present the density–magnetic field images at different evolutionary times in {Figure \ref{Fig2}}. These images are the slices along the Z-axis through the center of the box. The background is the density distribution, and the white arrows show the direction and intensity of the magnetic field. Both scenarios have a similar flow evolution. With the expansion of the ejecta, the forward shock sweeps up the ambient medium. In the northeast, the ejecta are considerably decelerated and mix with the high density ambient medium. We can find the swept gas piles up on the shock surface and the amplification of the magnetic field at the shock regions. 

\begin{figure}
\gridline{\fig{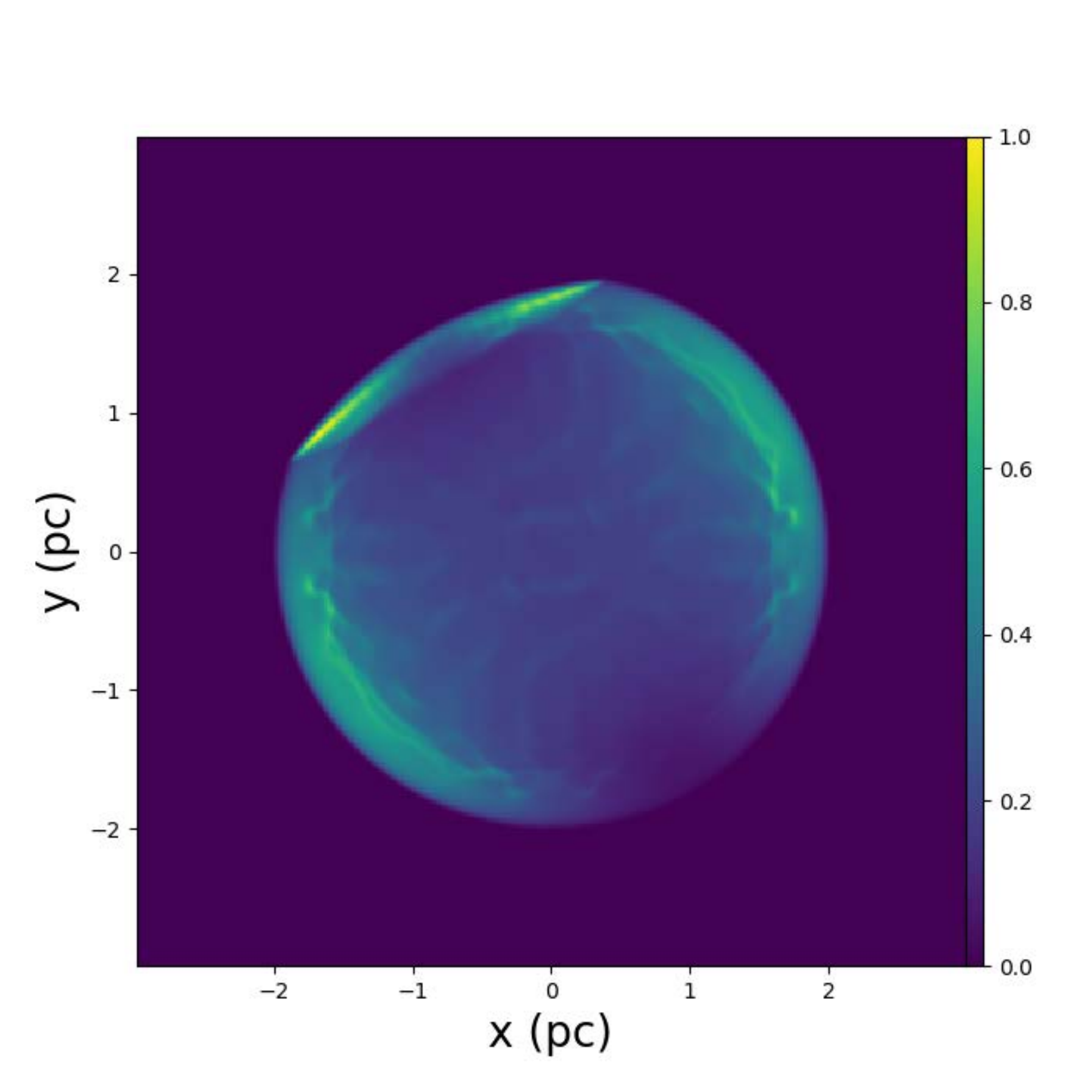}{0.49\textwidth}{(a)}
          \fig{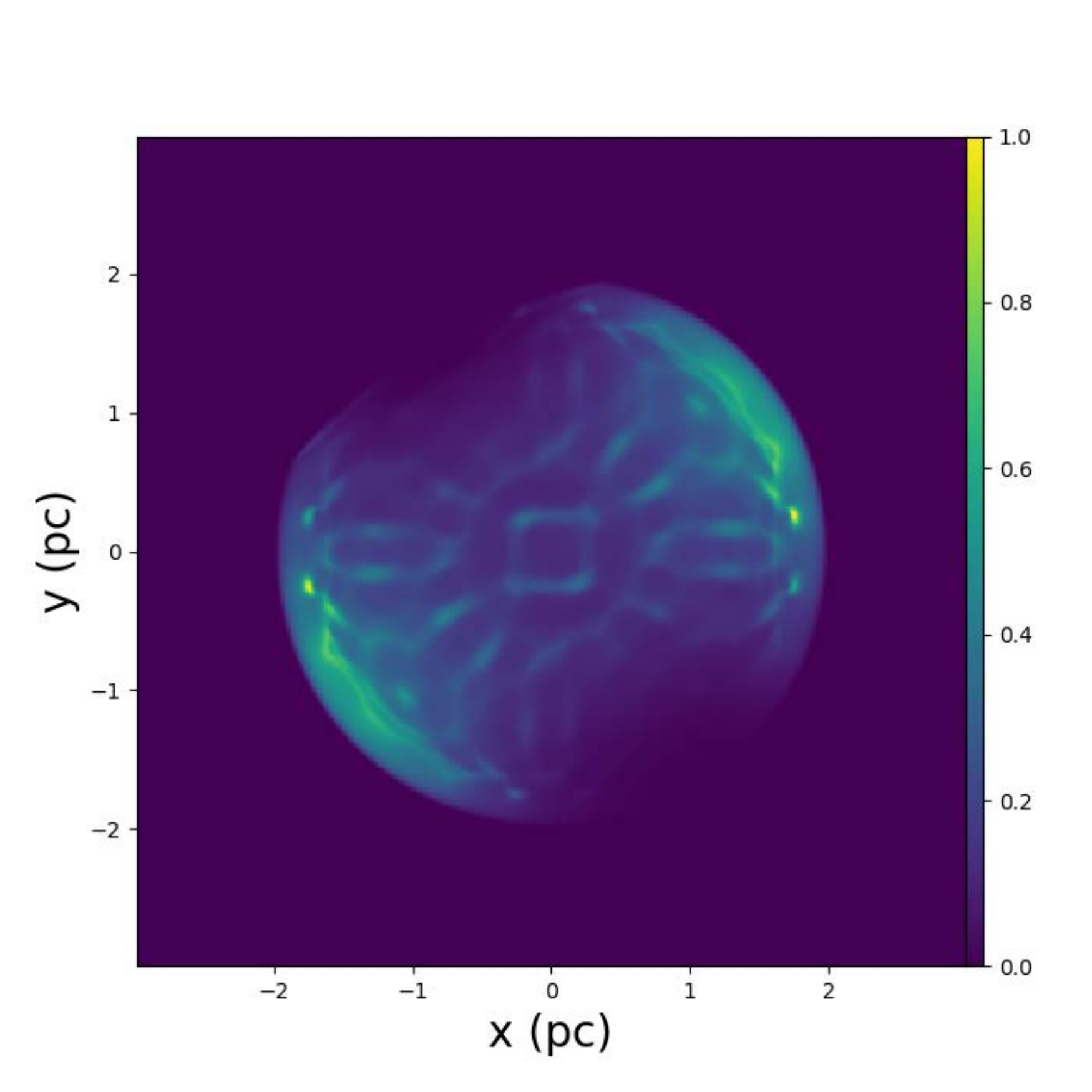}{0.49\textwidth}{(b)}}
\gridline{\fig{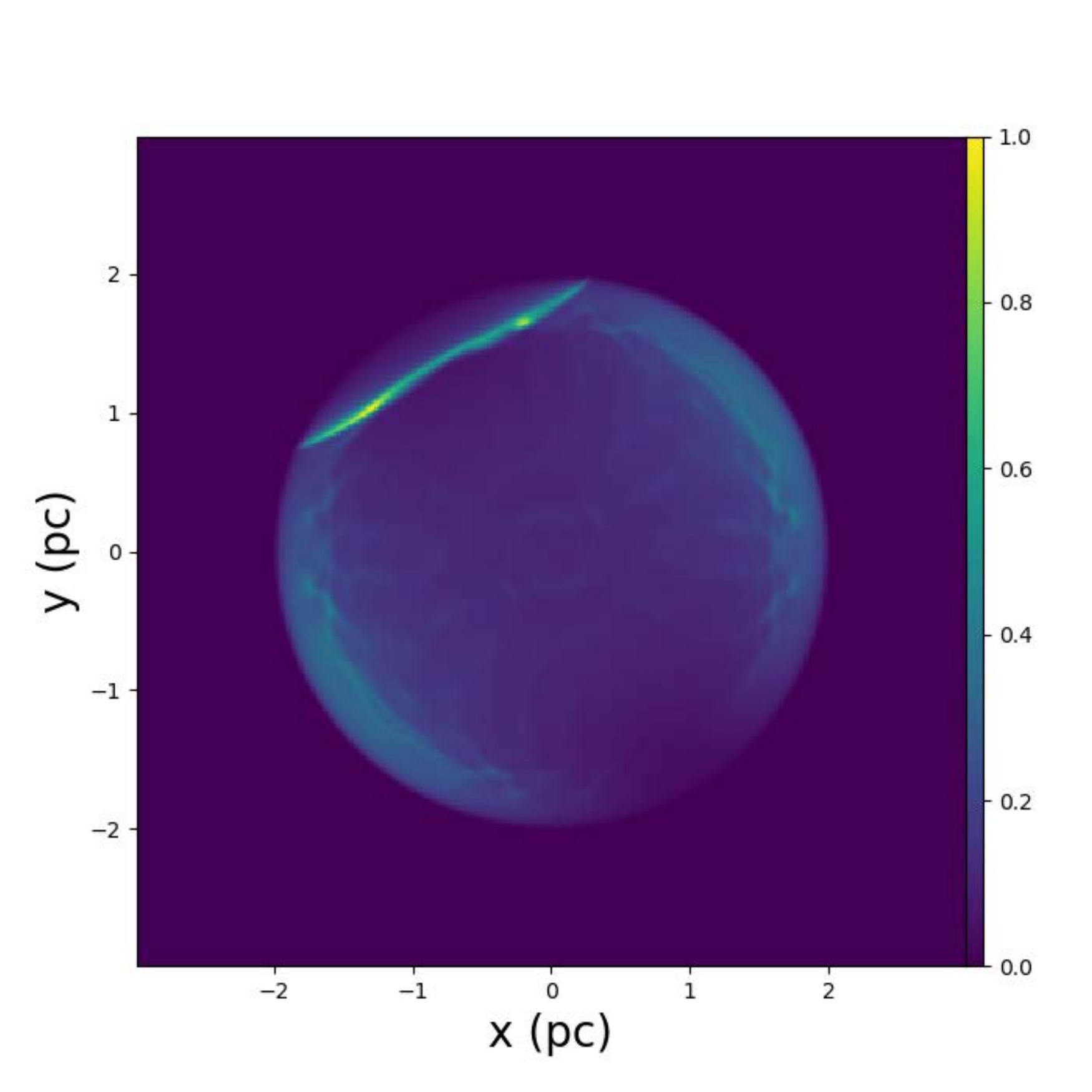}{0.49\textwidth}{(c)}
          \fig{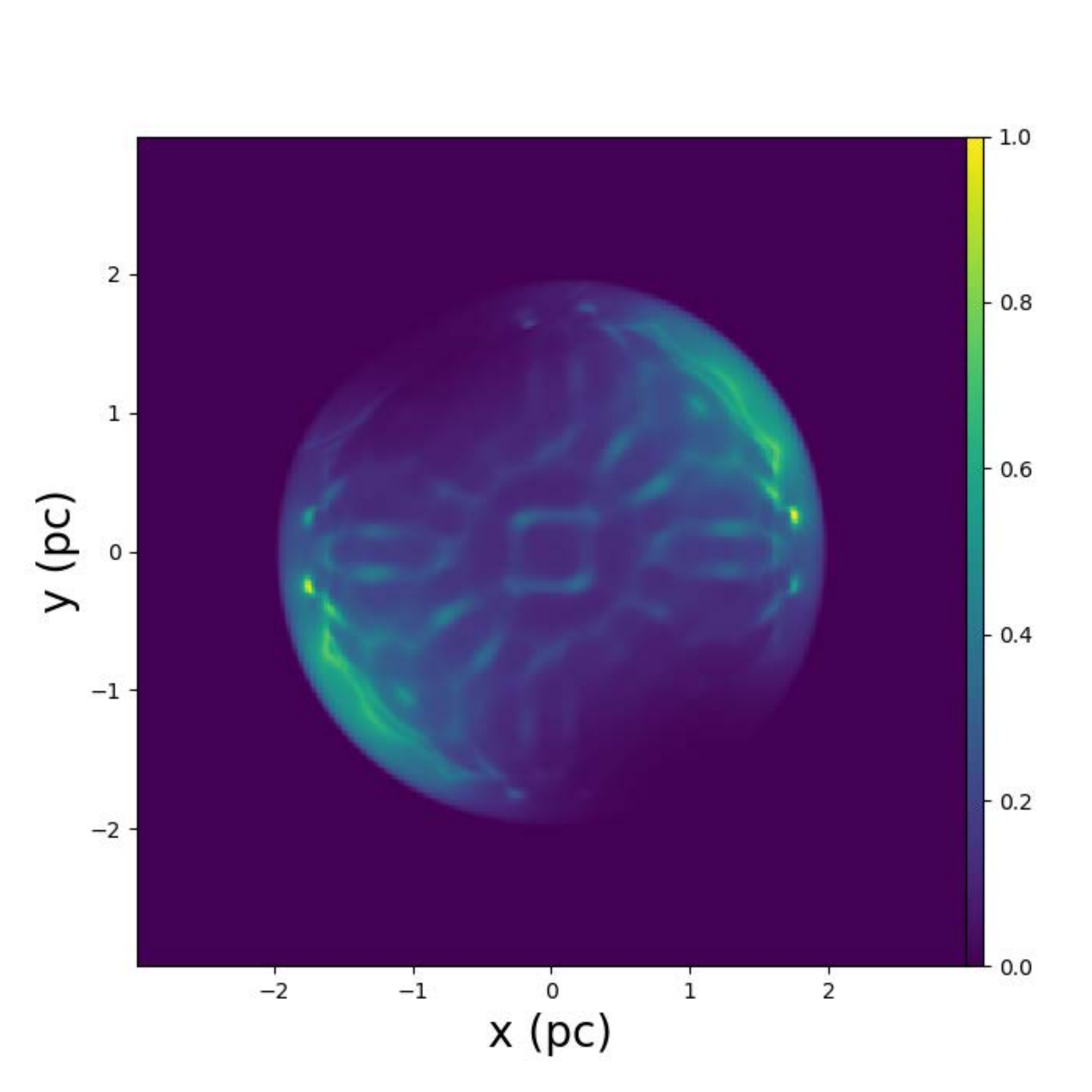}{0.49\textwidth}{(d)}}
\caption{{Simulated emission maps for different energy bands, with an age of 148 years. The normalized relative radiation flux distribution is showed in the figure. The top two pictures correspond to scenario A, and the bottom two pictures correspond to scenario B. Panels (a) and (c) show the radio image at 2.4 GHz, and panels (b) and (d)  show the X-ray image at 5 keV.}}
\label{Fig3}
\end{figure}

Assuming the synchrotron mechanism is dominant, we synthesize the relative emission flux in the radio and X-ray bands. We depict the projected emission map along the z-axis at 148 years in {Figure \ref{Fig3}}. Under both scenarios, the radio map shows an asymmetric shell with a bright east-north ring. While there is some difference in detail, the bright ring in the scenario of the clump is narrower than the scenario of the density gradient. In the X-ray band, the flux map for both scenarios features the symmetrical {limbs} on the east and west sides, and there are no bright structures in the northeast.

\section{Discussion} \label{discussion}
{As shown in Figure \ref{Fig3}, we tried to reproduce the morphology of SNR G1.9+0.3 in two assumed scenarios, and some features were reproduced in both cases. The differences between our simulations and observations are that some complex features are not reproduced, particularly the opposite 'ears' in the X-ray morphology. In this section, we shall discuss these weaknesses and limitations in our simulations.}

{'Ears' is a morphological feature often found in SNRs, which presents as a pair of antisymmetric protrusions on the outermost region of the remnant. For type Ia SNRs, \cite{2015MNRAS.450.1399T} derived a crude upper limit that 10–30 percent of them have opposite ear-like features. In general, the formation of 'ears' is attributed to the circumstellar medium or jets. The simulations by \cite{2021MNRAS.502..176C} showed that 'ears' can be produced in the case of an equatorially confined circumstellar structure. \cite{2022RAA....22l2003S} suggested 'ears' result from the type Ia supernova inside the planetary nebula, and jets can play a role in shaping 'ears'.}

{However, these peculiar structures are ignored in our simulations, which affects our reproduction of observed features. For simplification, we assume isotropic ejecta and neglect the progenitor's effect on the environment. For our case, we can roughly reproduce some observed features and also lose some of them. Our synthesized emission map shows two opposite arcs on the east and west sides, but no protrusion structures. Referring to \cite{2013MNRAS.435..320T}, we have also tried to add planetary nebulae and jets to our simulations, which show significant effects on the morphology of the remnant. These help to reproduce more complex features, but also enlarge the parameter space of the simulation. This issue is worth a further study.}

\section{Summary} \label{summary}

To investigate the formation of the discrete feature between the X-ray and radio morphology of SNR G1.9+0.3, we performed 3D MHD simulations and synthesized emission maps, which indicates it possibly originates from the joint effect of the density gradient, the dense gas clouds, and the magnetic field. 

Including the density gradient and the magnetic field, scenario A  shows, the shocks parallel to the magnetic field and flowing toward the denser region, will sweep through more interstellar matter and produce more non-thermal electrons, which contributes to the formation of the bright radio ring. \cite{2017ApJ...837L...7B} also implies the denser medium in the northeast based on the expansion measurements, but conclusive evidence still needs to be provided by higher-resolution observations. For type Ia supernova, the explosion energy and the ejecta mass can be well confined in a small parameter space, in which the simulation results will not largely change. 

The magnetic field distribution plays an important role in MHD simulations, but its initial configuration may not proper. In this work, we assume a uniform initial magnetic field and set the direction along the symmetry axis of the x-ray {arcs}. The evolved magnetic field and the synthesized X-ray map show similar symmetry features. The exponential cutoff of synchrotron emission occurs at frequencies higher than the critical frequency, and the critical frequency increases with the magnetic field strength for the same particle distribution.

As we have less knowledge about the ambient environment at the initial time, we are not aiming to reproduce the observed morphology one by one, but to find an acceptable explanation of the discrete feature.The simulated flux maps indicate that SNR G1.9+1.3 may have evolved in a similar environment. Further observations are needed if we want to impose tighter constraints on the parameter space of the remnant. More detailed simulations also require more discussion of processes such as particle acceleration, radiation mechanisms, etc.

\begin{acknowledgments}
This work is partially supported by National Key R\&D Program of China: 2018YFA0404203 \& 2018YFA0404202, by the National Natural Science Foundation of China (NSFC) under No.12073039 \& U1938112, and the China Manned Space Project with No. CMS-CSST-2021-A09. The computing task was carried out on the HPC cluster at China National Astronomical Data Center (NADC). NADC is a National Science and Technology Innovation Base hosted at National Astronomical Observatories, Chinese Academy of Sciences.
\end{acknowledgments}

%



\software{PLUTO \cite{2007ApJS..170..228M,2012ApJS..198....7M},}



\bibliography{sample631}{}
\bibliographystyle{aasjournal}



\end{document}